# Spatiotemporal Mapping of Photocurrent in a Monolayer Semiconductor Using a Diamond Quantum Sensor

Brian B. Zhou[1,2,*,†], Paul C. Jerger[1,*], Kan-Heng Lee[1,3], Masaya Fukami[1], Fauzia Mujid[4], Jiwoong Park[1,4], David D. Awschalom[1,5,†]

[1] Institute for Molecular Engineering, University of Chicago, Chicago, Illinois 60637, USA

[2] Department of Physics, Boston College, Chestnut Hill, Massachusetts 02467, USA

[3] School of Applied and Engineering Physics, Cornell University, Ithaca, NY 14853, USA

[4] Department of Chemistry and James Franck Institute, University of Chicago, Chicago, Illinois 60637, USA

[5] Institute for Molecular Engineering and Materials Science Division, Argonne National Laboratory, Lemont, Illinois 60439, USA

[*] These authors contributed equally to this work.

[†] email: brian.zhou@bc.edu, awsch@uchicago.edu

The detection of photocurrents is central to understanding and harnessing the interaction of light with matter. Although widely used, transport-based detection averages over spatial distributions and can suffer from low photocarrier collection efficiency. Here, we introduce a contact-free method to spatially resolve local photocurrent densities using a proximal quantum magnetometer. We interface monolayer $MoS_2$ with a near-surface ensemble of nitrogen-vacancy centers in diamond and map the generated photothermal current distribution through its magnetic field profile. By synchronizing the photoexcitation with dynamical decoupling of the sensor spin, we extend the sensor's quantum coherence and achieve sensitivities to alternating current densities as small as 20 nA/μm. Our spatiotemporal measurements reveal that the photocurrent circulates as vortices, manifesting the Nernst effect, and rises with a timescale indicative of the system's thermal properties. Our method establishes an unprecedented probe for optoelectronic phenomena, ideally suited to the emerging class of two-dimensional materials, and stimulates applications towards large-area photodetectors and stick-on sources of magnetic fields for quantum control.

## Introduction

The extraordinary features of two-dimensional van der Waals (2D vdW) systems have opened new directions for tailoring the interaction of light with matter, with potential to impact technologies for imaging, communications, and energy harvesting. The detection of photo-induced carriers is critical to realizing practical photosensing and photovoltaic devices[1–3], as well as to characterizing novel photo-responses, including optical manipulation of spin and valley indices[4,5], circular[6–8] and shift[9,10] photocurrents driven by non-trivial Berry curvature and scattering-protected photocurrents at a Dirac point[11]. Transport-based detection of photocurrents in 2D materials is susceptible to inefficient photocarrier extraction, requiring light to be directed near junctions with strong built-in electric field, and thus complicates the scale-up of devices to practical sizes[1–3]. To expand our understanding of light-matter interaction and overcome existing technical limitations, new photodetection approaches that offer high spatial and temporal resolution are needed.

In this work, we demonstrate a novel technique using an embedded quantum magnetometer[12–14] to detect and spatially resolve photocurrent densities via their local magnetic field signature. We transfer a monolayer $MoS_2$ (1$L$-$MoS_2$) sheet grown by metal-organic chemical vapor deposition[15] (MOCVD) onto a diamond chip hosting a near-surface ensemble of nitrogen-vacancy (NV) centers. The magnetic field due to photocurrents, driven in this case by the photothermoelectric effect (PTE)[16,17], modifies the quantum precession of the NV center spin, inducing a phase that can be optically detected[18–20]. Due to the near-field nature of our probe, our method does not require remote carrier extraction, and thus eliminates the need for electrical contacts and avoids challenges due to carrier trapping and potential fluctuations in large-area devices[2]. Moreover, in contrast to scanning photocurrent microscopy[6–11,16], our technique provides diffraction-limited spatial resolution for both excitation and detection (Fig. 1a). This enables detailed spatial information to be extracted even when the net photocurrent between two contacts in a conventional measurement is zero.

With wide phase space applicability and potential nanoscale spatial resolution, NV magnetometry has emerged as a premier tool for probing current distributions in materials[13], revealing insights on the structure of vortices in high-$T_c$ superconductors[21,22] and the effect of microscopic inhomogeneity on transport in graphene[23] and nanowires[24]. These demonstrations all probed direct current (dc) flow and were accordingly limited in sensitivity by the inhomogeneous dephasing time $T_2^*$ of the NV center. Here, we leverage the ability to control the timing of the photoexcitation to implement for the first time a "quantum lock-in" protocol to isolate alternating photocurrents. This protocol simultaneously decouples the NV center from wideband magnetic noise, extending its coherence time to the homogeneous $T_2$ limit, and achieves a sensitivity of 20 nA/μm for alternating current (ac) densities, fifty times smaller than previous work on dc current sensing in graphene (~1 μA/μm in Ref. [23]). Moreover, in contrast to dc sensing, our ac technique opens the investigation of the dynamics of photocarrier generation. Through changing the repetition rate of the photo-excitation pulses, we can resolve non-equilibrium response over a temporal bandwidth spanning from $1/T_2 \sim$ 10 kHz to an upper range of ~10 MHz, set by achievable driving speeds on the NV center spin[25].

I. **Hybrid NV-MoS$_2$ Photosensing Platform**

Figure 1a displays the experimental setup for detecting photocurrents in monolayer MoS$_2$. Two independently steerable laser beams (probe: 532 nm, excitation: 661 nm) are joined by dichroic mirrors and focused by a confocal microscope onto the MoS$_2$-diamond stack, held at a base temperature of 6 K (see Supplementary Section 1). We define the coordinate axes {x,y,z} to be parallel to the edges of the [001]-faced diamond sample with [110]-cut edges, as depicted in Fig. 1b. The position of the probe spot relative to the excitation spot on the sample is denoted by the vector ($R_x$, $R_y$). By aligning the external magnetic field $B_{ext}$ along the [111] axis, we selectively address a single subset out of the four possible NV lattice orientations, noting that full vector magnetometry can be achieved by extending our measurements to multiple orientations. An insulated wire coil placed over the MoS$_2$ monolayer delivers the resonant radio-frequency (RF) pulses for NV spin manipulation. Figure 1c diagrams the energy levels of the NV spin-triplet ground-state, showing the $|\pm 1\rangle$ sublevels separated from $|0\rangle$ by the zero-field splitting parameter $D = 2.87$ GHz. Applying $B_{ext}$ along the NV center axis lifts the degeneracy of the $|\pm 1\rangle$ states, which acquire equal and opposite Zeeman-shifts at a rate $\gamma = 2.8$ MHz/Gauss.

In Fig. 1d, we display an optical micrograph of a MoS$_2$-diamond stack assembled by our vacuum stacking technique[26]. The high-quality MOCVD-grown monolayer[15] readily covers our entire 2x2x0.5 mm diamond sample, but we deliberately expose a portion of the diamond to facilitate control measurements (see Supplementary Section 2). The slight absorption[27] of the 532 nm probe laser by the single atomic layer of MoS$_2$ does not interfere with initialization of the NV center spin state. However, monolayer MoS$_2$'s strong intrinsic photoluminescence (PL) from 532 nm excitation partially overlaps the NV emission spectrum and overwhelms the signal of single NV centers, precluding their identification upon coverage (Fig. 1e). To increase the NV signal and facilitate arbitrary spatial mapping, we instead utilize an engineered diamond sample hosting an ensemble of near-surface NV centers (~40 nm deep; ~85 NV centers per focused optical spot). Additionally, we band-pass filter the detected PL between 690 nm and 830 nm to predominantly isolate NV center emission, as shown in the room-temperature PL spectra of monolayer MoS$_2$ and a typical ensemble NV sample (Fig. 1f). Importantly, the excitation wavelength for photocarriers in MoS$_2$ must be sufficiently longer than the zero-phonon line of the NV center (637 nm) to minimize thermally-assisted absorption[28] by the NV center during the photocurrent sensing duration. Excitation and subsequent decay of the NV center will decohere the spin superposition, leading to reduced read-out contrast and sensitivity[29]. We excite at 661 nm, but have verified that our effects persist for longer excitation wavelength (see Supplementary Section 5.1).

Our photocurrent sensing protocol is based on an XY8-*N* dynamical decoupling sequence commonly used for the NV-based detection of ac fields from the precession of remote nuclei[18–20]. However, in contrast to applications in nuclear magnetic resonance, here we directly control both the frequency and the phase of the targeted oscillating field through the timing of the photo-excitation pulses (Fig. 2a). This enables us to sweep the phase of the oscillating field relative to the NV sensing sequence and examine the full dependence, increasing sensitivity compared to measuring an averaged response over random phases. We prepare the superposition state $|\psi_i\rangle = 1/\sqrt{2}\,(|0\rangle + |-1\rangle)$ and allow it to evolve to $|\psi_f\rangle = 1/\sqrt{2}\,(|0\rangle + e^{i\phi}|-1\rangle)$ under the influence of photo-excitation and the XY8-*N* rephasing pulses (*N* repetitions of 8 $\pi$-pulses). We first match the spacing $\tau$ between the $\pi$-pulses to a half period of the photo-excitation frequency $\nu_{exc}$ (= $1/2\tau$ here) and probe the acquired phase $\phi$ for varying relative delays $\theta$ between the start of the $\pi$-pulses and the photo-excitation (Fig. 2a). The *X* and *Y*-projections $X_P = \cos(\phi)$ and $Y_P = \sin(\phi)$, respectively, of the final state $|\psi_f\rangle$ are optically detected after an appropriate projection pulse on the NV center.

In essence, the delay $\theta$ of our "quantum lock-in" protocol plays the analogous role to the relative phase between the signal and reference oscillator in a classical lock-in measurement. If photoexcitation generates an instantaneous, square pulse current density $\vec{J}$ in the MoS$_2$ monolayer, then the phase accumulated by the NV center will be maximized for an optimal delay $\theta_{opt} = 0°$. We denote this maximal accumulated phase as $\Phi$, with its amplitude and sign determined by the amplitude and direction of the local current density. Alternatively, if the photocurrent rises and falls with a characteristic timescale (purple trace in Fig. 2a), maximum phase accumulation will occur for nonzero optimal delay ($\theta_{opt} > 0°$) (see Supplementary Section 4.1). For a current density $\vec{J}$ with sinusoidal time-dependence, the accumulated phase $\phi$ will depend on $\theta$ as $\phi = \Phi \cos(\theta - \theta_{opt})$. This form represents a good approximation to our data due to the smoothing effect of the photocurrent rise and fall times. In Fig. 2b, we plot the analytical behaviors for $X_P$ and $Y_P$ under this model as a function of the delay $\theta$ and the maximal phase $\Phi$.

## II. Detection and Mapping of Photo-Nernst Currents

We first perform a photocurrent sensing protocol with $N = 2$ and $\tau = 7.6$ $\mu$s over an uncovered area of diamond (probe and excitation beams slightly offset). Consistent with negligible absorption by the NV center or bulk diamond at 661 nm, we cannot detect the presence of photoexcitation and measure $\phi = 0$ for all $\theta$ (Supplementary Fig. S6). Remarkably, when we shift to an area where monolayer MoS$_2$ covers the diamond, we detect oscillations in $X_P$ and $Y_P$ as $\theta$ is varied (Fig. 2c; $R_X = 1.07$ $\mu$m, $R_Y = 0$ $\mu$m), revealing the presence of an ac magnetic field due to photocurrents. Fitting the $X_P$ and $Y_P$ projections simultaneously to their expected behavior (solid lines), we deduce that the maximal phase $\Phi$ increases in magnitude as the photoexcitation power $P$ increases and that the optimized delay $\theta_{opt}$ is nonzero, indicating a finite photocurrent rise time $\tau_{rise}$ (Fig. 2c). In principle, MoS$_2$-assisted heating of the diamond substrate could induce a time-periodic change in the zero-field splitting $D$ and also lead to effective precession via detuning of our microwave pulses from resonance. However, we rule out this scenario by showing that the phase accumulated by the initial state $|\psi_i\rangle = 1/\sqrt{2}\,(|0\rangle + |+1\rangle)$ is opposite to that by $1/\sqrt{2}\,(|0\rangle + |-1\rangle)$, while changes in $D$ are expected to affect both $|\pm 1\rangle$ symmetrically (Supplementary Fig. S7).

Strikingly, when the probe beam is moved to the opposite side of excitation beam ($R_X = -0.67$ $\mu$m, $R_Y = 0$ $\mu$m), the phase accumulated by the NV center switches sign, signifying a reversal in direction of the local photocurrent density $\vec{J}$ (Fig. 2d). For both locations, we summarize the dependence of the maximum phase $\Phi$ on the photoexcitation power, showing a sublinear increase (Fig. 2e). This behavior is consistent with a PTE origin for the photocurrents as the thermal gradient induced in monolayer MoS$_2$ by laser heating begins to saturate based on our thermal modeling (Supplementary Section 6). The absence of any interface potentials that induce directional electric fields in our implementation further suggests a PTE origin, which has been shown to be the dominant photocurrent generation mechanism in monolayer MoS$_2$ in prior experiments[16,17]. Finally, by increasing the sensing duration through the number of repetitions $N$ of the XY8 block, we can increase $\Phi$ linearly (Fig. 2f).

The maximum accumulated phase $\Phi$ represents a weighted time-integral of the ac magnetic field along the NV axis produced by the photocurrents. By modeling the pulse shape of $\vec{J}$, we can estimate the final instantaneous field $B_{max}$ from the time-integrated field (Fig. 2a) via:

$$B_{max} = \beta * \Phi / (0.5 * 8N\tau * 2\pi\gamma),$$

where $\beta$ is a pulse shape dependent factor, $\Phi$ is measured in radians, and the factor of 0.5 stems from the 50% duty-cycle of the photo-excitation. The factor $\beta$ increases monotonically with $\tau_{rise}$ from $\beta = 1$ for square pulses to $\beta = 2$ as $\tau_{rise} \to \infty$ (see Supplementary Section 4.1). Unless otherwise stated, we utilize a constant factor $\beta = 1.25$, corresponding roughly to the range of our typical measurements ($\tau$ = 7.6 μs). In Fig. 2e, we resolve $B_{max}$ as small as $0.84 \pm 0.08$ mG for about 2 hours of averaging time ($0.14 \pm 0.05$ mG for additional data shown in Supplementary Fig. S8). From this, we assert a minimum sensitivity to a sheet current density of ~20 nA/μm flowing perpendicular to the NV axis (along the y-direction), which produces a field $B_{max}$ ~ 0.1 mG independent of the depth of the NV center. This improvement in sensitivity compared to dc current sensing (~1 μA/μm in graphene[23]) is enabled by our synchronized dynamical decoupling protocol, which extends the NV ensemble's coherence time from $T_2^* = 0.51\ \mu s$ to $T_2(XY8-2) = 235\ \mu s$ and provides access to coherent oscillations in $Y_P = \sin(\phi)$, more sensitive than $X_P = \cos(\phi)$ to small $\phi$.

In steady state, the divergence-free condition $\nabla \cdot \vec{J} = 0$ and rotational symmetry of our experiment imply that any photocurrent must flow as vortices around the excitation spot, explaining the reversal in the direction of $\vec{J}$ observed in Fig. 2c and d. Nonzero photocurrent, defining a chirality to the vortex, also requires the breaking of time-reversal symmetry, supplied here by the external magnetic field. We deem the resulting current profile a "photo-Nernst vortex", since the radial temperature gradient induced by the excitation beam and the out-of-plane $B_{ext}$ result in azimuthal current flow, everywhere transverse to the temperature gradient as in the Nernst effect. Previously, photo-Nernst currents[30,31] were detected by scanning photocurrent microscopy at the edges of exfoliated graphene devices. However, spatial mapping of an unperturbed vortex in the interior of a 2D material has not been possible, as notably it generates zero net current in a transport measurement.

By scanning the probe beam relative to the excitation spot, we map the magnetic field distribution generated by the photo-Nernst vortex. In Fig. 3a and b, we present the measured $B_{max}$ for line scans along the x- and y-axes, respectively. Crucially, we show that $B_{max}$ changes sign as expected when the static magnetic field *B* is reversed, indicating that the chirality of the vortex also reverses (Fig. 3a). The solid lines in Fig. 3a and b present the simulated field profiles at the NV center depth for a model of the photocurrent distribution $\vec{J}(R)$. We assume an azimuthal flow with amplitude

$$|\vec{J}(R)| \propto \frac{d}{dR} e^{-\frac{R^2}{2\sigma_J^2}}$$

since $\vec{J}(R)$ is expected to be proportional to the gradient of an approximately Gaussian photo-induced temperature distribution (Supplementary Section 7). Although we phenomenologically incorporate deviations from a perfectly circular excitation beam to better match the experimental profile, the salient features of a vortex current density are clear. $B_{max}$ is nonzero at the vortex center due to the z-component of the field produced by the current loops. Cutting across the vortex along x, the fringing fields beneath the plane of the current loops either align or anti-align with the oblique [111]-oriented NV axis, leading to the asymmetry in the magnitude of $B_{max}$ for positive versus negative $R_X$ (Fig. 3a). For the y-direction, $B_{max}$ becomes slightly negative as we pass outside the ring of maximum current density and the z-component of the fringing field reverses (Fig. 3b).

In Fig. 3c, we plot the current distribution $\vec{J}(R)$ used to approximate the experimental field profile together with independent thermal modeling of the laser-induced temperature distribution $T_M(R)$ in monolayer MoS$_2$. The modeled $\vec{J}(R)$ peaks at $\sigma_J \sim 1.0~\mu m$, in close agreement with the predicted location of the maximum thermal gradient (Supplementary Fig. S15). At our base temperature for the diamond substrate ($T_D$ = 6 K), the photocurrent vortex is enhanced by the reduced thermal conductivity of monolayer MoS$_2$ and the large thermal interface resistance to the substrate, which permit large thermal gradients (~18 K/$\mu m$ max) and a spatial distribution significantly larger than the excitation spot size (Gaussian intensity with standard deviation $\sigma_{exc} = 0.45~\mu m$). As $T_D$ increases, the thermal conductivity[32] and thermal interface conductance[33] both increase and we find that the detected $B_{max}$ at $R_X = -0.95~\mu m$ diminishes, disappearing around 20 K (Supplementary Fig. S10).

Using $\vec{J}(R)$, we estimate that the integrated current for one side of the vortex is ~1.3 µA for an excitation power of 25 uW before the objective (85% transmission). This implies a Nernst photo-responsivity of ~60 mA/W for 226 G parallel to the NV axis (130 G perpendicular to the sample). For the same magnetic field, this value for ungated monolayer MoS$_2$ is higher than the giant Nernst photo-responsivities reported for a graphene-hexagonal boron nitride heterostructure that is gate-tuned to its van Hove singularities[31]. This enhancement in MoS$_2$ is consistent with its lower thermal conductivity and higher Seebeck coefficient stemming from a favorable density of states for its gapped band structure[16,17]. In Supplementary Fig. S10, we verify that the Nernst photocurrent is linear in the external magnetic field $B_{ext}$ and non-saturating up to 500 G, as expected for the low field regime[31].

Our unique probe provides additional insight into the dynamics of photocarrier generation. In Fig. 3d, we examine the optimal delay $\theta_{opt}$ between the NV driving and photoexcitation pulses as a function of $R_X$, using a sequence with $\tau$ = 7.6 $\mu$s. As the probe beam moves away from the excitation spot, $\theta_{opt}$ increases. This effect can be explained if the rise time for the local photocurrent, which dominates the contribution to the local field, increases for larger $|R_X|$. To corroborate this hypothesis, we map the leading edge of the photocurrent rise by varying the pulse spacing $\tau$ in the synchronized sensing protocol. To deduce $B_{max}$, the value of the field at the end of the pulse, we need to account for variations in pulse shape as $\tau$ changes. For each set with different $\tau$, we utilize the measured delay $\theta_{opt}$ to infer the factor $\beta$ within our pulse shape model. In Fig. 3e, we compare $B_{max}(R,\tau)$ for two different locations. We confirm an exponential rise to the photocurrent with a time constant $\tau_{rise} \sim 1$ $\mu$s that increases for larger $|R_X|$. The extracted rise times are sufficient to explain the measured $\theta_{opt}$, suggesting that no additional effects, such as carrier propagation from the excitation spot, contribute significantly to the delay (see Supplementary Section 5.5). Indeed, we find that $\tau_{rise}$ is, to within error, independent of the external magnetic field, which would affect carrier propagation.

The rise times for $B_{max}$ can be compared to a model of system's transient thermal response. Indeed, the rise time for the thermal gradient $dT_M(R)/dR$ increases for larger $|R|$ (Supplementary Section 6). This matches the qualitative experimental trend (Fig. 3c,d) and supports a picture of photocurrents generated locally by PTE. Interestingly, to approximate the microsecond-scale photocurrent rise times, we need to assume a heat capacity $c_p$ for monolayer $MoS_2$ that is significantly higher than theoretically predicted[34,35]. Our model estimates $c_p \sim 200$ J/(kg $\ast K^2$) $\ast T_M$ for temperatures below ~50 K, while $c_p$ is generally taken[16] as 400 J/(kg $\ast$ K) for single crystal monolayer $MoS_2$ at 300 K. Even considering that we use polycrystalline $MoS_2$, this discrepancy may suggest extrinsic contributions to the estimated $c_p$. For example, excess heat capacity could arise from PMMA residue or a layer of cryopumped adsorbates, and the latter is known to significantly raise the measured low-temperature heat capacity of other low-dimensional materials[36,37] (see Supplementary Fig. S2). Further investigations under systematic outgassing and sample cleaning conditions are required to clarify this phenomena, as well as to explore potential applications toward the sensing of absorbed gases.

Finally, we demonstrate the ability to detect light without prior knowledge of its frequency or phase. Gating the light at a constant frequency $\nu_{exc}$ with an independent controller, we examine the projection $X_P$ of the final state $|\psi_f\rangle$ as we scan the spacing $\tau$ of the XY8-*8* sequence. When the frequency $\nu = 1/2\tau$ of the decoupling sequence matches $\nu_{exc}$ to within a bandwidth $\Delta\nu \approx .11/N\tau$, the average value of $X_P$ over random starting delays $\theta$ is diminished from its initial full projection, resulting in a resonant dip[19]. In Fig. 4, we demonstrate this unsynchronized detection scheme for three different frequencies $\nu_{exc}$ = 65 kHz, 110 kHz, and 333 kHz. Due to the rise time of the PTE photocurrents, our sensitivity to optical power decreases for higher $\nu_{exc}$, necessitating stronger excitation to see the same contrast change. However, the NV sensing protocol itself is effective for frequencies up to several tens of MHz[25] and thus can be combined with faster photocurrent mechanisms for optimal photodetection.

**Discussion**

Our demonstration broaches wide-ranging opportunities for investigating fascinating opto-electronic phenomena in materials. Highlighted by our mapping of a photo-Nernst vortex, the functionality of spatial resolution lacking in transport-based techniques could allow clear-cut characterization of chiral photocurrents localized at edges[38], directional currents controlled by coherent optical injection[39], and the scattering of valley-polarized[4], Weyl point[8–10], or Dirac-point[11] photocurrents from disorder. Beyond fundamental interests, nano-engineering of the diamond surface into pillared arrays would reduce the thermal interface conductance and possibly extend the results here to room temperature. This, in conjunction with the use of wide-field imaging techniques[23] and isotopically purified samples[40] that prolong the NV's quantum coherence, could enable ultrasensitive, large-area photodetector arrays using our technique. Moreover, the optical generation of ac magnetic fields that are spatially localized, but also available on-demand anywhere across the area covered by $MoS_2$ provides simplified, stick-on alternatives for fabricating devices to manipulate solid-state spins. Adding to capabilities such as electrical spin readout[41], spontaneous emission tuning[42] and 2D ferromagnetism[43], our demonstration of spin-photocurrent interaction widens the perspective for integrated quantum technologies based on the quantum emitter-2D material platform.

**References**

1.  Koppens, F. H. L. *et al.* Photodetectors based on graphene, other two-dimensional materials and hybrid systems. *Nat. Nanotech.* **9,** 780–793 (2014).

2. Mak, K. F. & Shan, J. Photonics and optoelectronics of 2D semiconductor transition metal dichalcogenides. *Nat. Photon.* **10,** 216–226 (2016).

3. Liu, Y. *et al.* Van der Waals heterostructures and devices. *Nat. Rev. Mater.* **490,** 16042 (2016).

4. Mak, K. F., McGill, K. L., Park, J. & McEuen, P. L. The valley Hall effect in MoS2 transistors. *Science* **344,** 1489–1492 (2014).

5. Ye, Z., Sun, D. & Heinz, T. F. Optical manipulation of valley pseudospin. *Nat. Phys.* **13,** 26–29 (2016).

6. Yuan, H. *et al.* Generation and electric control of spin–valley-coupled circular photogalvanic current in WSe2. *Nat. Nanotech.* **9,** 851–857 (2014).

7. Eginligil, M. *et al.* Dichroic spin–valley photocurrent in monolayer molybdenum disulphide. *Nat. Commun.* **6,** 7636 (2015).

8. Ma, Q. *et al.* Direct optical detection of Weyl fermion chirality in a topological semimetal. *Nat. Phys.* **13,** 842–847 (2017).

9. Osterhoudt, G. B. *et al.* Colossal mid-infrared bulk photovoltaic effect in a type-I Weyl semimetal. *Nat. Mater.* (2019). doi:10.1038/s41563-019-0297-4

10. Ma, J. *et al.* Nonlinear photoresponse of type-II Weyl semimetals. *Nat. Mater.* (2019). doi:10.1038/s41563-019-0296-5

11. Ma, Q. *et al.* Giant intrinsic photoresponse in pristine graphene. *Nat. Nanotech.* **14,** 145–150 (2019).

12. Degen, C. L., Reinhard, F. & Cappellaro, P. Quantum sensing. *Rev. Mod. Phys.* **89,** 035002 (2017).

13. Casola, F., van der Sar, T. & Yacoby, A. Probing condensed matter physics with magnetometry based on nitrogen-vacancy centres in diamond. *Nat. Rev. Mater.* **3,** 17088 (2018).

14. Awschalom, D. D., Hanson, R., Wrachtrup, J. & Zhou, B. B. Quantum technologies with optically interfaced solid-state spins. *Nat. Photon.* **12,** 516–527 (2018).

15. Kang, K. *et al.* High-mobility three-atom-thick semiconducting films with wafer-scale homogeneity. *Nature* **520,** 656–660 (2015).

16. Buscema, M. *et al.* Large and Tunable Photothermoelectric Effect in Single-Layer MoS 2. *Nano Lett.* **13,** 358–363 (2013).

17. Hippalgaonkar, K. *et al.* High thermoelectric power factor in two-dimensional crystals of MoS2. *Phys. Rev. B* **95,** 115407 (2017).

18. Staudacher, T. *et al.* Nuclear magnetic resonance spectroscopy on a (5-nanometer)3 sample volume. *Science* **339,** 561–563 (2013).

19. DeVience, S. J. *et al.* Nanoscale NMR spectroscopy and imaging of multiple nuclear species. *Nat. Nanotech.* **10,** 129–134 (2015).

20. Lovchinsky, I. *et al.* Magnetic resonance spectroscopy of an atomically thin material using a single-spin qubit. *Science* **355,** 503–507 (2017).

21. Thiel, L. *et al.* Quantitative nanoscale vortex imaging using a cryogenic quantum magnetometer. *Nat. Nanotech.* **11,** 677–681 (2016).

22. Pelliccione, M. *et al.* Scanned probe imaging of nanoscale magnetism at cryogenic temperatures with a single-spin quantum sensor. *Nat. Nanotech.* **11,** 700–705 (2016).

23. Tetienne, J. *et al.* Quantum imaging of current flow in graphene. *Sci. Adv.* **3,** e1602429 (2017).

24. Chang, K., Eichler, A., Rhensius, J., Lorenzelli, L. & Degen, C. L. Nanoscale Imaging of Current Density with a Single-Spin Magnetometer. *Nano Lett.* **17,** 2367–2373 (2017).

25. Romach, Y. *et al.* Spectroscopy of Surface-Induced Noise Using Shallow Spins in Diamond. *Phys. Rev. Lett.* **114,** 017601 (2015).

26. Kang, K. *et al.* Layer-by-layer assembly of two-dimensional materials into wafer-scale heterostructures. *Nature* **550,** 229–233 (2017).

27. Bae, J. J. *et al.* Thickness-dependent in-plane thermal conductivity of suspended MoS 2 grown by chemical vapor deposition. *Nanoscale* **9,** 2541–2547 (2017).

28. Kern, M. *et al.* Optical cryocooling of diamond. *Phys. Rev. B* **95,** 235306 (2017).

29. Fuchs, G. D., Falk, A. L., Dobrovitski, V. V. & Awschalom, D. D. Spin Coherence during Optical Excitation of a Single Nitrogen-Vacancy Center in Diamond. *Phys. Rev. Lett.* **108,** 157602 (2012).

30. Cao, H. *et al.* Photo-Nernst current in graphene. *Nat. Phys.* **12,** 236–239 (2016).

31. Wu, S. *et al.* Multiple hot-carrier collection in photo-excited graphene Moiré superlattices. *Sci. Adv.* **2,** e1600002 (2016).

32. Yarali, M. *et al.* Effects of Defects on the Temperature-Dependent Thermal Conductivity of Suspended Monolayer Molybdenum Disulfide Grown by Chemical Vapor Deposition. *Adv. Funct. Mater.* **27,** 1704357 (2017).

33. Ong, Z.-Y., Cai, Y. & Zhang, G. Theory of substrate-directed heat dissipation for single-layer graphene and other two-dimensional crystals. *Phys. Rev. B* **94,** 165427 (2016).

34. Su, J., Liu, Z., Feng, L. & Li, N. Effect of temperature on thermal properties of monolayer $MoS_2$ sheet. *J. Alloys Compd.* **622,** 777–782 (2015).

35. Saha, D. & Mahapatra, S. Analytical insight into the lattice thermal conductivity and heat capacity of monolayer $MoS_2$. *Phys. E* **83,** 455–460 (2016).

36. Hone, J., Batlogg, B., Benes, Z., Johnson, A. T. & Fischer, J. E. Quantized Phonon Spectrum of Single-Wall Carbon Nanotubes. *Science* **289,** 1730–1733 (2000).

37. Lasjaunias, J. C., Biljaković, K., Benes, Z., Fischer, J. E. & Monceau, P. Low-temperature specific heat of single-wall carbon nanotubes. *Phys. Rev. B* **65,** 113409 (2002).

38. Karch, J. *et al.* Terahertz Radiation Driven Chiral Edge Currents in Graphene. *Phys. Rev. Lett.* **107,** 276601 (2011).

39. Sun, D. *et al.* Coherent control of ballistic photocurrents in multilayer epitaxial graphene using quantum interference. *Nano Lett.* **10,** 1293–1296 (2010).

40. Balasubramanian, G. *et al.* Ultralong spin coherence time in isotopically engineered diamond. *Nat. Mater.* **8,** 383–387 (2009).

41. Brenneis, A. *et al.* Ultrafast electronic readout of diamond nitrogen–vacancy centres coupled to graphene. *Nat. Nanotech.* **10,** 135–139 (2014).

42. Tielrooij, K. J. *et al.* Electrical control of optical emitter relaxation pathways enabled by graphene. *Nat. Phys.* **11,** 281–287 (2015).

43. Thiel, L. *et al.* Probing magnetism in 2D materials at the nanoscale with single spin microscopy. *Preprint at* arXiv:1902.01406v1 (2019). doi:arXiv:1902.01406v1

**Methods**

*Sample Fabrication*

An NV center ensemble was created ~40 nm deep into an [001]-oriented diamond sample by $^{15}N^+$ ion implantation. The implantation energy was 30 keV, with an area dose of $10^{12}$ ions/cm$^2$. The samples were annealed at 850 °C and then 1100 °C to form roughly 85 NVs per optical spot. The external field $B_{ext}$ was supplied by a permanent magnet, and microwave pulses were delivered by a wire coil suspended above the sample surface.

Monolayer MoS$_2$ was grown on SiO$_2$/Si via MOCVD. It was then spin-coated with PMMA and baked at 180 °C. After applying thermal release tape (TRT), the stack was mechanically peeled off the SiO$_2$/Si substrate and transferred to the diamond in a vacuum chamber. Lastly, the TRT and PMMA were removed. Three separate MoS$_2$ samples were transferred and investigated in this work, all displaying NV-based photocurrent detection.

Further details of the sample growth and fabrication can be found in the Supplementary Information.

*NV Photocurrent Sensing Technique*

The photocurrent sensing sequence consists of simultaneous ac optical excitation of the monolayer MoS$_2$ and dynamical decoupling pulses applied to the NV center spin. A 532 nm laser for NV spin readout and 661 nm laser for photocurrent excitation were focused by an objective onto the sample held at 6 K inside a closed-cycle cryostat (Montana Instruments). The 661 nm laser spot was laterally displaced by impinging on the objective's back aperture at a slight angle away from normal. The displacement was measured by collecting reflected light off the sample into a camera and determining the center locations of the beams using calibrated pixel sizes. The 661 nm excitation was pulsed by modulating its polarization with an electro-optical modulator and passing the output through a Glan-Thompson polarizer. The polarization of the 661 nm beam was thereafter set to be right circularly polarized; however, the results here are not dependent on the polarization.

After optically initializing the NV spin into |0⟩, the XY8-*N* dynamical decoupling sequence applied to the NV center consists of 8*N*+2 qubit rotations:

$$\left(\frac{\pi}{2}\right)_y - \left[\pi_y - \pi_x - \pi_y - \pi_x - \pi_x - \pi_y - \pi_x - \pi_y\right]^N - \left(\frac{\pi}{2}\right)_{proj}$$

Here, the subscript indicates the axis on the Bloch sphere for the qubit rotation, and $\{\pi/2, \pi\}$ indicates the rotation angle. The axis of the final $\pi/2$ projection pulse determines which component $\{X, Y\}$ of the final spin superposition is rotated onto the $|0\rangle$ (bright) state for readout. The $\pi$ pulses are uniformly spaced by the interval $\tau$, whereas the $\pi/2$ pulses are spaced by $\tau/2$. The timing (frequency, delay) of the photoexcitation and NV microwave pulses can be synchronously controlled to nanosecond resolution using an arbitrary waveform generator.

*Data Analysis and Modeling*

All errorbars reported in our paper are 95% confidence intervals. Complete details of the data fitting, thermal modeling and stray field modeling are supplied in the Supplementary Information.

**Acknowlegments**

We thank C. F. de las Casas, A. L. Yeats, C. P. Anderson for experimental suggestions, and K. Burch for illuminating discussions. B.B.Z., P.C.J., M.F., and D.D.A. were supported by AFOSR FA9550-14-1-0231, ARO MURI W911NF-14-1-0016, ONR N00014-17-1-3026, and the University of Chicago MRSEC DMR-1420709. B.B.Z also acknowleges support from Boston College startup funding. K.H.L, F.M., J.P. were supported by the AFOSR (FA9550-16-1-0031, FA9550-18-1-0480), and the NSF through MRSEC programs at Cornell (DMR-1719875) and the University of Chicago (DMR-1420709), and the Materials Innovation Platform program (DMR-1539918).

**Author Contributions**

B.B.Z conceived and demonstrated the NV photosensing technique. P.C.J and B.B.Z performed NV center experiments. M.F. measured photoluminescence spectra. K.H.L fabricated the hybrid monolayer MoS$_2$–diamond samples, using MoS$_2$ synthesized by F.M. D.D.A and J.P. coordinated the collaboration. All authors contributed to the data analysis and manuscript preparation.

The authors declare no competing financial interests. Correspondence and requests for materials should be addressed to B.B.Z. or D.D.A.

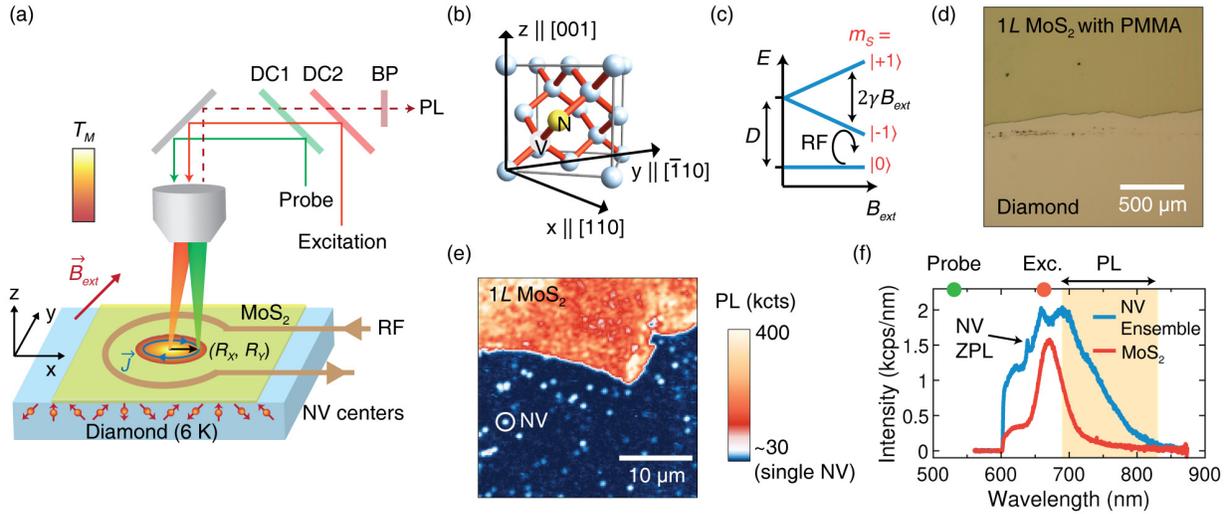

**Figure 1 | Photosensing platform based on monolayer MoS$_2$ and NV centers**

a) Experimental setup. A MOCVD-grown monolayer MoS$_2$ sheet is transferred on top of a bulk diamond sample hosting a near-surface ensemble of NV centers. Illuminating the MoS$_2$ with an excitation laser (661 nm) generates a temperature distribution $T_M$ within the monolayer that drives a photocurrent distribution $\vec{J}$. The NV center spin state senses the local magnetic field produced by the photocurrents and is optically read-out by a separate green laser (532 nm). DC1 – dichroic mirror (550 nm); DC2 – dichroic mirror (685 nm); BP – bandpass (690-830 nm); PL – photoluminescence. b) Crystal structure of the NV center in the diamond lattice. We address the subset of NV centers aligned with the [111]-direction and define the coordinate directions *x* and *y* as indicated. c) Energy levels of the NV ground-state triplet as a function of $B_{ext}$, the magnetic field parallel to the NV center axis. Resonant microwave pulses (RF) shown prepare and manipulate an equal superposition of the $m_s = |0\rangle$ and $|-1\rangle$ states for phase acquisition. d) Optical micrograph of monolayer MoS$_2$ on top of a diamond substrate after vacuum transfer. To enhance optical contrast, this micrograph is taken prior to cleaning off the poly(methyl methacrylate) (PMMA) coating layer, which supports the MoS$_2$ during transfer. e) Room-temperature PL image of the boundary region between monolayer MoS$_2$ and a bare diamond substrate, containing single NV centers. Image taken with DC1 and DC2 filters only. f) Room-temperature PL spectrum of monolayer MoS$_2$ and an ensemble NV sample under 532 nm illumination. The photoexcitation wavelength (661 nm) used in the sensing sequence is longer than the NV zero-phonon line (ZPL; 637 nm) to minimize optical excitation of the NV center. The orange shaded region (690-830 nm) depicts the region of collected PL for NV spin readout.

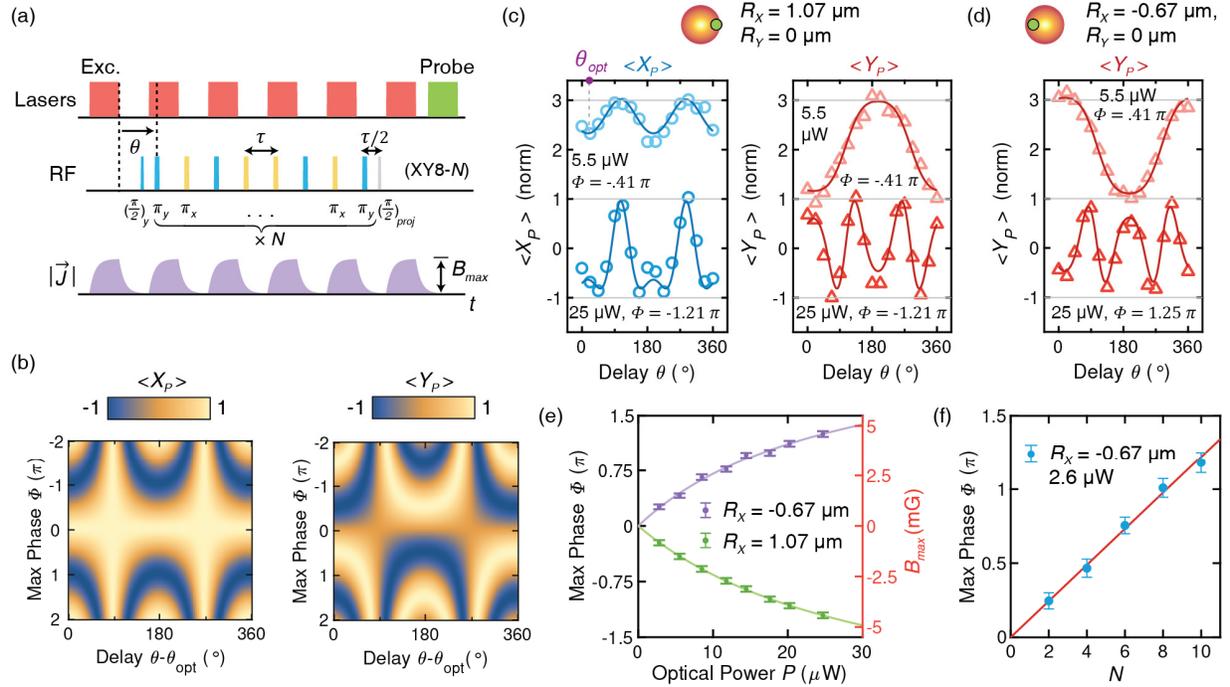

**Figure 2 | Dynamically-decoupled sensing protocol for detection of ac photocurrents**

a) Synchronized photo-excitation and dynamical decoupling sequence. The excitation laser is gated at a frequency $\nu_{exc} = 1/2\tau$, while an XY8-$N$ dynamical decoupling sequence with spacing $\tau$ between the $\pi$-pulses is applied to the NV center spin. The delay between the first $\pi$-pulse and the end of the last photoexcitation pulse is denoted as $\theta$, which is given as a phase (e.g., $\theta = 225°$ for the pulse trains shown). The phase of the last RF $\pi/2$ projection pulse prior to readout by the probe laser determines which projection ($X$ or $Y$) of the final NV center superposition state is measured. Bottom: schematic of the time-dependent current density generated in monolayer MoS$_2$. b) Expected dependence of the final NV spin projections $X_P$ and $Y_P$ on the delay $\theta$ (horizontal axis) and the maximum phase $\Phi$ acquired at the optimal delay $\theta_{opt}$ (vertical axis). The sign and magnitude of $\Phi$ are determined by the direction and magnitude of the local photocurrents. c) Experimental $X_P$ and $Y_P$ projections of the final NV spin state as the delay $\theta$ is varied for two different optical powers (sequence with $N = 2$, $\tau = 7.6~\mu s$). The probe beam ($R_X = 1.07~\mu m$) is positioned to the right of the excitation beam. The solid lines are simultaneous fits to both projections, allowing determination of $\Phi$ and $\theta_{opt}$. d) Same measurement but with the probe beam ($R_X = -0.67~\mu m$) to the left of the excitation beam. $X_P$ (not shown) looks similar to c), but $Y_P$ is inverted, indicating that the local photocurrent direction is reversed. e) Dependence of the maximum phase $\Phi$ on the optical power $P$ for $R_X = -0.67~\mu m$ (purple) and $R_X = 1.07~\mu m$ (green).

The right-hand axis converts $\Phi$ to the maximal field amplitude of the pulse, $B_{max}$, along the NV axis. f) Dependence of $\Phi$ on the number of repetitions $N$ of the XY8-$N$ sequence ($\tau$ = 7.6 $\mu$s).

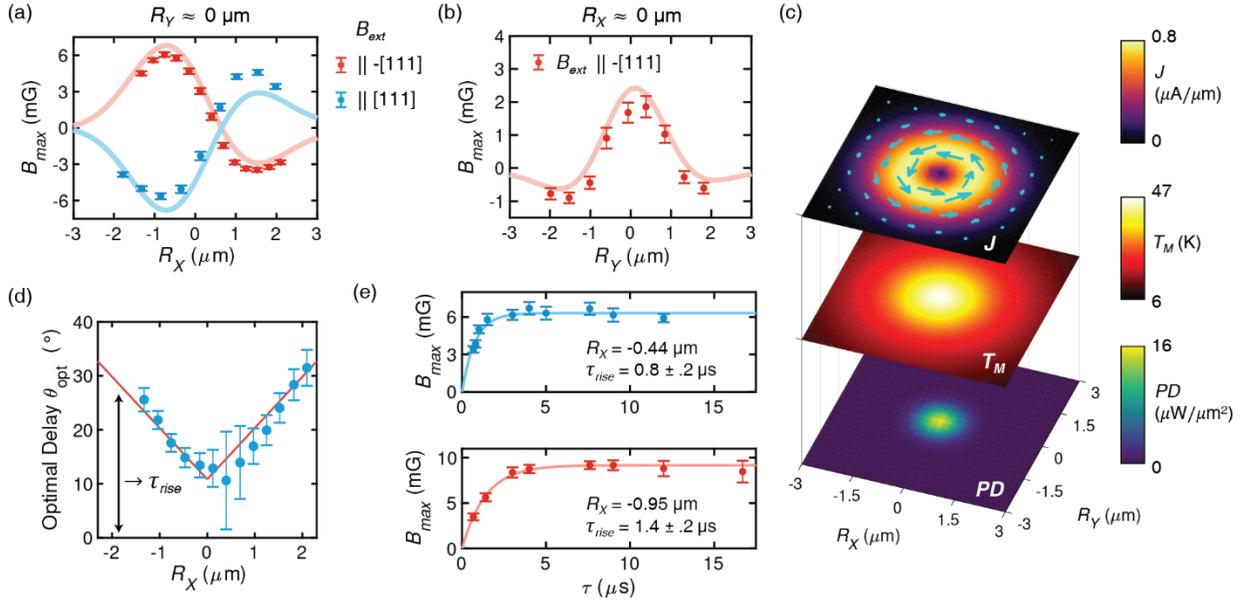

**Figure 3 | Spatiotemporal mapping of a photo-Nernst vortex**

a) Line scan of the maximal stray field $B_{max}$ as the probe moves along the x-axis of the vortex ($R_Y \approx 0\ \mu m$, $P = 25\ \mu W$). $B_{max}$ is extracted from the phase acquired while sweeping the delay $\theta$. When the direction of the external magnetic field $B_{ext}$ is reversed, $B_{max}$ changes sign in accordance with the Nernst effect (here we fix the positive field direction to be along [111]). b) Line scan of $B_{max}$ across the y-axis ($R_X \approx 0\ \mu m$). For the y-axis, the stray fields of the photocurrent vortex have a strong component perpendicular to the NV-axis, and $B_{max}$ is dominated by the z-component of the total field, which diminishes away from the vortex center. The solid lines in a) and b) are simulated stray fields along the NV axis for a modeled vortex current density $\vec{J}(R)$; slightly different parameters are used for a) and b). c) Comparison of the modeled current density $\vec{J}(R)$, the simulated laser-induced temperature distribution $T_M(R)$ in the monolayer MoS$_2$, and the power density $PD(R)$ of the excitation beam for $P = 25\ \mu W$, assuming rotational symmetry and using the measured excitation beamwidth. The spatial profile of $\vec{J}(R)$ is in close agreement with the gradient of $T_M(R)$. d) Dependence of the optimal delay $\theta_{opt}$ on the coordinate $R_X$. The solid line is a linear fit. e) Measured $B_{max}$ versus the spacing $\tau$ of the synchronized photosensing sequence for two different values of $R_X$ using $P = 38\ \mu W$. Here, the timing of the photoexcitation and NV driving pulses are changed together. The rise time $\tau_{rise}$ for $B_{max}$ increases for larger $|R_X|$, corroborating the trend in $\theta_{opt}$ shown in d). The solid lines are exponential fits.

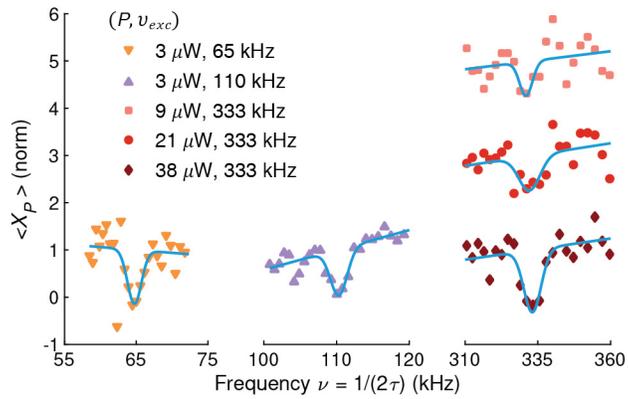

**Figure 4 | Detection of unsynchronized, variable frequency ac photocurrents**

The *X* projection of the final NV state using an unsynchronized XY8-8 sensing sequence. The photoexcitation frequency $\nu_{exc}$ is fixed for each set and the frequency of the NV driving pulses $\nu = 1/2\tau$ is scanned. When $\nu$ approaches resonance with $\nu_{exc}$, the projection $X_P$ displays a dip when averaged over repetitions with random relative phase between the photoexcitation and NV driving pulses. Data for three different photoexcitation frequencies $\nu_{exc}$ = 65 kHz, 110 kHz and 333 kHz are shown. The amplitude of the resonant dip is sensitive to the photoexcitation power $P$.